# Predicting and extending the bubble-free operating limit of capillary-fed water electrolysis

*Farkhondeh Khodabandeh*[1], *Zhengmao Lu*[1,*]

[1] Institute of Mechanical Engineering, EPFL, 1015 Lausanne, Switzerland

[*] Lead contact and correspondence: Zhengmao Lu (zhengmao.lu@epfl.ch)

**Summary**

Gas bubbles impose large penalties in water electrolysis at industrially relevant rates. Capillary-fed electrolyzers greatly mitigate them by wicking electrolyte to electrodes and venting gas from the backside, but only below a threshold current density. We show that the separator pore size sets this threshold through two opposing effects: smaller pores raise the supersaturation required for nucleation but reduce the permeability for electrolyte replenishment. A coupled electrochemical-transport model, anchored to a single measured onset, quantifies both mechanisms across 0.1–8 μm pores. Experimentally, reducing pore size from 8 to 1.2 μm raises the bubble onset limit from 0.18 to 1.73 A $cm^{-2}$ and lowers cell voltage by 0.11 V at 1 A $cm^{-2}$, primarily by suppressing bubble-induced overpotential. At 0.45 μm and below, the replenishment limit takes over and the cells dry out before bubbles form. Our findings establish the bubble-free operation limit as designable rather than purely empirical.

## Introduction

Green hydrogen from water electrolysis links renewable electricity to decarbonization of energy-intensive sectors.[1,2] Its economic and energy viability depends on electrolyzer cell efficiency at industrially relevant current densities,[3,4] precisely where the two-phase environment created by continuous gas evolution imposes the greatest penalties.[5,6] Gas bubbles cloak catalytic surfaces, impede electrolyte access, and distort ion conduction pathways, imposing mass transport and ohmic losses that grow as current density rises.[7–11] Evacuating gas from electrodes while delivering electrolyte to them is therefore the central design problem in high-rate electrolysis.[12–16]

Capillary-fed electrolysis (CFE)[17] disentangles competing transport demands architecturally: electrolyte is drawn from a remote reservoir through a porous separator to the catalyst layers, while hydrogen and oxygen are released directly into separate backside gas chambers (Figure 1). With no bulk liquid surrounding the electrodes, evolved gas leaves without traversing the ionic pathway, and bubble coverage of the catalyst is largely suppressed. As a result, CFE has delivered cell voltages well below those of incumbent technologies, including commercial alkaline and proton exchange membrane electrolyzers, at comparable current densities.[17] Subsequent acoustic detection at room temperature, however, showed that bubble-free operation persisted only up to approximately 0.17 A $cm^{-2}$, beyond which bubbles emerged and imposed an additional voltage penalty.[18] That some limit exists is expected: dissolved gas concentration rises with current density until supersaturation triggers nucleation. Nevertheless, what sets that threshold, and whether it can be engineered, remains poorly understood.

Conventional bubble management strategies act after nucleation, promoting detachment through forced flow,[19,20] wettability,[21,22] topography engineering,[23,24] pulsed operation,[25,26] or applied fields;[27,28] in a capillary-fed cell the question is whether a bubble forms at all. Here, we identify the separator pore size as the key design parameter. Dissolved gas accumulates in the thin electrolyte layer wetting the electrode,[29,30] and supersaturation is the highest at the boundary between the catalyst layer and the separator, where the separator pores set the geometry available to a forming nucleus. The onset supersaturation scales inversely with the pore size,[31–34] which corresponds to a nucleation-limited bubble-free current density through dissolved gas transport. On the other hand, finer pores reduce electrolyte transport capacity, giving rise to a lower dryout current density, above which water consumption outpaces capillary replenishment. The two limits oppose one another through the same parameter, and the maximum bubble-free current density is set by whichever is lower. To determine the resulting optimum requires a model that predicts both limits quantitatively, which has not been established for CFE, leaving the trade-off unresolved.

In this study, we develop such a model and use it to predict and extend the bubble-free limit of capillary-fed alkaline electrolysis. A coupled transport-based and electrochemical framework resolves capillary electrolyte delivery, dissolved gas transport, ionic conduction, and the polarization response, simultaneously. From a single reference cell, we extract the electrode kinetic parameters at current densities below 5 mA cm$^{-2}$, where the response is activation-controlled and free of bubbles. Holding these parameters fixed, together with independently determined transport properties, the model then yields the complete bubble-free polarization baseline. For the reference cell, experimentally measured voltages depart from this baseline at a current density that coincides with the independently reported, acoustically measured onset,[18] validating the approach. For nonreference cells, comparison of the deviation current density with the predicted nucleation and capillary limits identifies the controlling mechanism.

Extending this analysis across pore sizes of 0.1, 0.45, 1.2, 5, and 8 μm, we determine whether bubble nucleation or capillary replenishment controls the operating limit. The nucleation criterion, anchored to the single reference onset, predicts the controlling mechanism without further adjustment. Notably, reducing the pore size from 8 to 1.2 μm raises the bubble-free limit nearly tenfold, and reduces the bubble-induced overpotential to a negligible level at 1 A cm$^{-2}$. Below 1.2 μm, the benefit of suppressing nucleation was progressively offset by the loss of permeability. This trade-off became evident for the 0.45 μm separator and dominant for the 0.1 μm separator, where electrolyte replenishment could no longer sustain the imposed current. Bounded by nucleation at large pore sizes and by dryout at small ones, the bubble-free limit is thus maximized at an intermediate pore size, largest for the 1.2 μm separator among those tested, reaching 1.73 A cm$^{-2}$. More broadly, our results recast the bubble-free limit from an empirical ceiling into a quantity that can be predicted from pore sizes and extended by design.

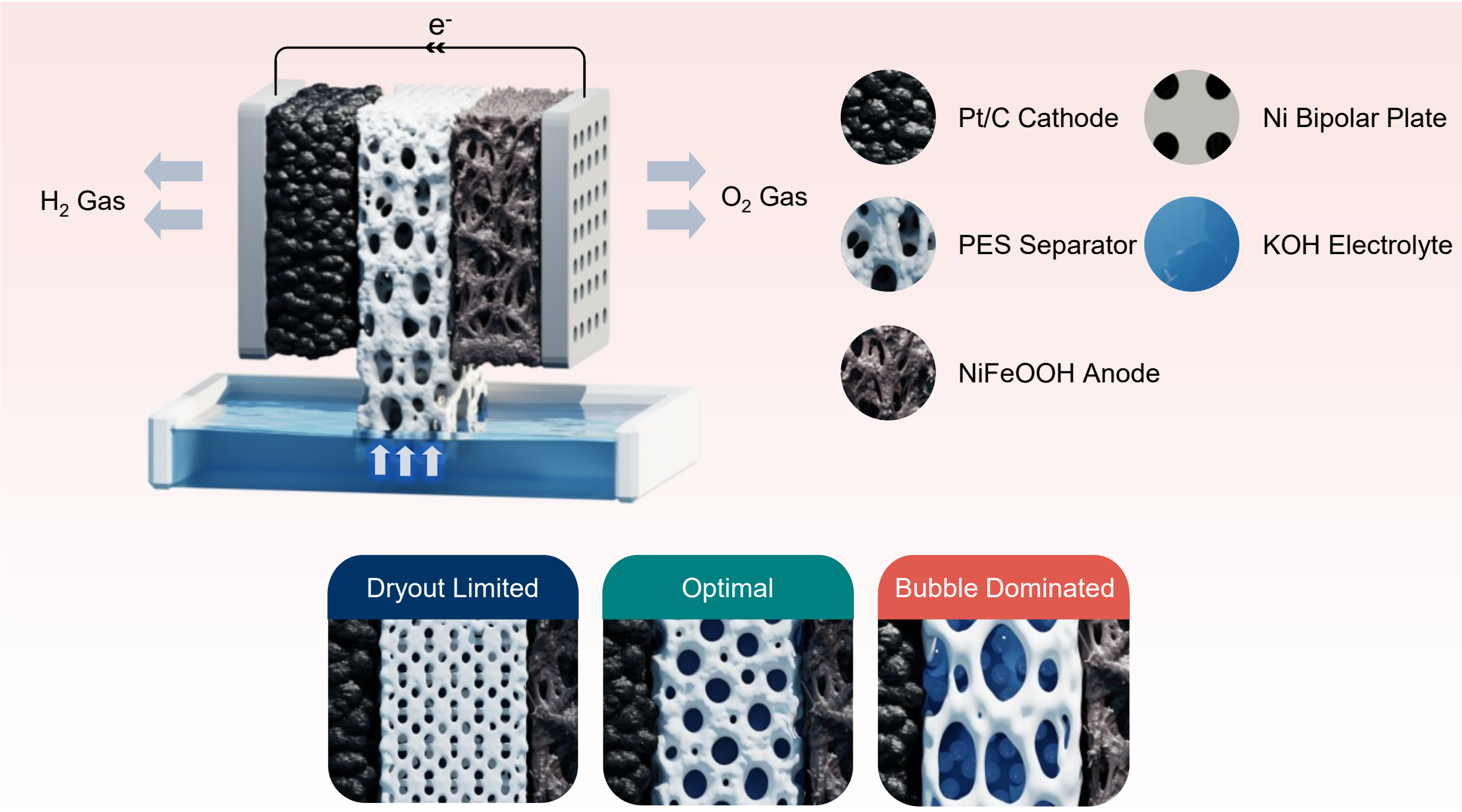

**Figure 1. Separator pore size governs the balance between bubble suppression and electrolyte transport in capillary-fed water electrolysis.** KOH electrolyte is drawn from a remote reservoir through the PES separator to the Pt/C cathode and NiFeOOH anode, while $H_2$ and $O_2$ are released into separate gas chambers. Large separator pores promote bubble formation and the associated voltage penalty, whereas excessively small pores restrict electrolyte replenishment and induce transport limitation. An intermediate pore size balances these competing effects, enabling optimal bubble-free operation. The illustration is schematic and not to scale.

## Results

### Model Framework

Prediction of the CFE operating limit requires resolving three coupled physical processes: bubble-free polarization, dissolved gas transport, and capillary electrolyte replenishment, each described by its own set of governing equations.

*Bubble-free polarization model*

To isolate the voltage penalty introduced by bubbles, we construct a bubble-free polarization baseline by resolving activation, concentration, and ohmic losses that set the cell voltage in the absence of a bubble phase. For each electrode, we solve for ionic transport and the potential distribution on a one-dimensional domain spanning from the catalyst surface, through the electrical double layer (EDL), to the bulk electrolyte in the pore space (Figure 2A,B). Following the Gouy-Chapman-Stern description of the EDL,[35,36] each domain comprises a Stern layer between the electrode and the outer Helmholtz plane (OHP), followed by the electrolyte region from the OHP to the bulk. The Stern layer is treated as a compact capacitive region with a linear potential drop. This drop represents the activation overpotential $\eta_{\mathrm{act},i}$ and is prescribed by Tafel kinetics:[37,38]

$$\eta_{\mathrm{act},i} = \left|\Phi_{\mathrm{e},i} - \Phi_{\mathrm{OHP},i}\right| = \eta_{0,i} \log_{10}\left(\frac{j}{j_{0,i}}\right) = -\lambda_{\mathrm{s},i}\mathbf{n} \cdot [\nabla\Phi]_{\mathrm{OHP},i}\,, \qquad i = \mathrm{c}, \mathrm{a}. \tag{1}$$

where $\Phi_{\mathrm{OHP},i}$ and $\Phi_{\mathrm{e},i}$ are the respective potentials at the OHP and the solid surface, and $i$ = c, a denotes the cathode and anode domains, respectively. $\eta_{0,i}$ is the Tafel slope, $j$ is current density, $j_{0,i}$ is the exchange current density, $\lambda_{\mathrm{s},i}$ is the Stern layer thickness and $\mathbf{n}$ is the surface normal vector. Beyond the OHP, the concentration overpotential $\eta_{\mathrm{conc},i}$ is obtained by solving the generalized modified Poisson-Nernst-Planck (GMPNP) equations.[16,39,40] The total molar flux $\boldsymbol{J}_n$ includes diffusion, migration, and finite-ion-size effects:

$$\boldsymbol{J}_n = -D_n\nabla C_n - \frac{z_n F}{RT} D_n C_n \nabla\Phi - D_n C_n \left(\frac{N_\mathrm{A}\sum_n a_n^3\,\nabla C_n}{1 - N_\mathrm{A}\sum_n a_n^3\,C_n}\right), \qquad n = \mathrm{OH^-}, \mathrm{K^+}, \mathrm{H^+}. \tag{2}$$

where $D_n$ is the diffusion coefficient, $C_n$ is the molar concentration, $z_n$ is the ionic charge number, $F$ is the Faraday constant, $R$ is the gas constant, $T$ is the temperature, $\Phi$ is the potential field, $N_\mathrm{A}$ is the Avogadro constant and $a_n$ is the effective solvated diameter of species $n$. At steady state, conservation of each ionic species requires

$$\nabla \cdot \boldsymbol{J}_n = 0. \tag{3}$$

The potential field is governed by Poisson's equation,

$$\nabla \cdot (\varepsilon_0 \varepsilon_\mathrm{r} \nabla\Phi) = -F \sum_n z_n C_n, \tag{4}$$

where $\varepsilon_0$ is the vacuum permittivity and $\varepsilon_\mathrm{r}$ is the relative permittivity of the electrolyte. The fixed double layer and transport parameters are provided in Table S1. At the OHP, $K^+$ and $H^+$ have zero flux, while $OH^-$ is produced at the cathode and consumed at the anode

$$\boldsymbol{J}_{\mathrm{OH^-,c}} = \frac{j}{F}\,, \qquad \boldsymbol{J}_{\mathrm{OH^-,a}} = -\frac{j}{F}. \tag{5}$$

At the bulk boundary, $C_n$ = $C_{\mathrm{bulk},n}$ and $\Phi$ is set to be 0 such that the concentration overpotential $\eta_{\mathrm{conc},i} = |\Phi_{\mathrm{OHP},i}|$.

The resulting bubble-free cell voltage $V_{\mathrm{bf}}$ is

$$V_{\mathrm{bf}} = E_{\mathrm{rev}} + \sum_{i=\mathrm{a,c}} (\eta_{\mathrm{act},i} + \eta_{\mathrm{conc},i}) + jR_{\mathrm{s}}, \tag{6}$$

where $E_{\mathrm{rev}}$ = 1.23 V at room temperature and $R_s$ is the measured area-specific ohmic series resistance. For operation dominated by bubbles, the difference between the experimentally measured cell voltage ($V_{\mathrm{exp}}$) and the predicted bubble-free voltage defines the bubble-induced overpotential:

$$\eta_{\mathrm{b}}(j) = V_{\mathrm{exp}}(j) - V_{\mathrm{bf}}(j). \tag{7}$$

The current density at which the experimental and bubble-free voltages first satisfy the deviation criterion is denoted as $j_{\mathrm{dev}}$. Depending on the controlling regime, $j_{\mathrm{dev}}$ indicates either bubble emergence or the onset of capillary limitation. The kinetic parameter fitting and quantitative determination of $j_{\mathrm{dev}}$ are described in Methods.

*Dissolved gas transport and bubble-onset model*

Dissolved $H_2$ and $O_2$ transport is modeled in a two-dimensional cross-section of the cathode-separator-anode stack (Figure 2C). The separator is treated as liquid-filled, while electrolyte penetration into each electrode is represented by an effective wetted layer of thickness $t_{\mathrm{f}}$ adjacent to the separator. Water consumption, water generation, and gas evolution through electrochemical reactions are confined to these layers. The dissolved gas flux prior to nucleation is predicted by

$$\boldsymbol{J}_g = -D_{\mathrm{eff},g,i} \nabla C_g + C_g \boldsymbol{u}, \qquad g = \mathrm{H_2, O_2}, \qquad i = \mathrm{c, s, a}. \tag{8}$$

where $\boldsymbol{J}_g$ and $C_g$ are the molar flux and concentration of gas $g$. The evaluation of the effective gas diffusivity $D_{\mathrm{eff},g,i}$ is specified in Methods and the sensitivity analysis of the wetted-layer thickness can be found in Note S1. Here, $i$ is the domain index, with $i$ = c, s, a corresponding to the cathode, separator, and anode, respectively. The electrolyte velocity $\boldsymbol{u}$ is approximated by the Darcy velocity obtained from steady, incompressible liquid transport:

$$\nabla \cdot (\rho_{\mathrm{KOH}} \boldsymbol{u}) = Q_{\mathrm{m},i}, \qquad \boldsymbol{u} = -\frac{\kappa_i}{\mu_{\mathrm{KOH}}} (\nabla p_{\mathrm{l},i} - \rho_{\mathrm{KOH}} \boldsymbol{g}), \tag{9}$$

where $\rho_{\mathrm{KOH}}$ and $\mu_{\mathrm{KOH}}$ are the electrolyte density and viscosity, respectively, $p_{\mathrm{l},i}$ is the liquid pressure, $\boldsymbol{g}$ is the gravitational acceleration, and $Q_{\mathrm{m},i}$ is the mass source associated with electrochemical reactions. The permeability $\kappa_i$ and the uniformly distributed electrochemical source terms are defined in Methods. The pressure at the separator/reservoir boundary is fixed at the atmospheric pressure $p_{\mathrm{r}} = p_{\mathrm{atm}}$, while the remaining external boundaries have zero liquid flux. Steady-state conservation of dissolved gas is given by

$$\nabla \cdot \boldsymbol{J}_g = R_{g,i}, \tag{10}$$

where $R_{g,i}$ is the volumetric gas generation rate, confined to the corresponding wetted layer; the gas-specific source terms are described in Methods. At the separator-reservoir boundary, the gas concentration is fixed at its bulk value $C_{\text{bulk},g}$, while at the boundaries facing the gas chambers, gas-liquid equilibrium is imposed using Henry's law:

$$C_g = K_{\text{H},g} p_{\text{atm},g}, \tag{11}$$

where $K_{\text{H},g}$ is the Henry's solubility constant and $p_{\text{atm},g}$ is the partial pressure of gas $g$ in the corresponding chamber. All remaining external boundaries have zero gas flux.

Bubble nucleation requires the dissolved gas pressure to overcome the capillary pressure resisting bubble formation within the separator pore. Combining Henry's law with the inverse dependence of capillary pressure on pore size gives the dimensionless nucleation factor

$$NF_g = \frac{C_{\max,g} d_{\text{s}}}{K_{\text{H},g} \gamma_{\text{KOH}}}, \tag{12}$$

where $C_{\max,g}$ is the maximum dissolved gas concentration before nucleation, $d_{\text{s}}$ is the nominal separator pore diameter, and $\gamma_{\text{KOH}}$ is the electrolyte surface tension. For each gas, $NF_g$ is determined from the predicted $C_{\max,g}$ at the experimentally measured bubble onset current density of the 8 μm separator. Assuming $NF_g$ is independent of separator pore size, the maximum concentration required for nucleation is calculated for each separator pore geometry. For each nonreference separator, the nucleation current density, $j_{\text{nuc}}$, is defined as the lower of the current densities at which the predicted hydrogen and oxygen concentrations reach their respective nucleation values. Bubble onset is therefore evaluated independently through the experimentally determined polarization curve deviation, $j_{\text{dev}}$, and the gas transport model prediction, $j_{\text{nuc}}$.

*Capillary replenishment and dryout model*

At sufficiently high current density, electrochemical water consumption can exceed the maximum electrolyte flow sustained by capillary suction, defining the capillary-limited current density, $j_{\text{cap}}$. Its calculation accounts for capillary pressure, hydrostatic head, viscous losses, and distributed water withdrawal along the separator; the complete derivation is provided in Note S2. The bubble-free operating limit, $j_{\text{lim}}$, is therefore set by the lower of $j_{\text{nuc}}$ and $j_{\text{cap}}$.

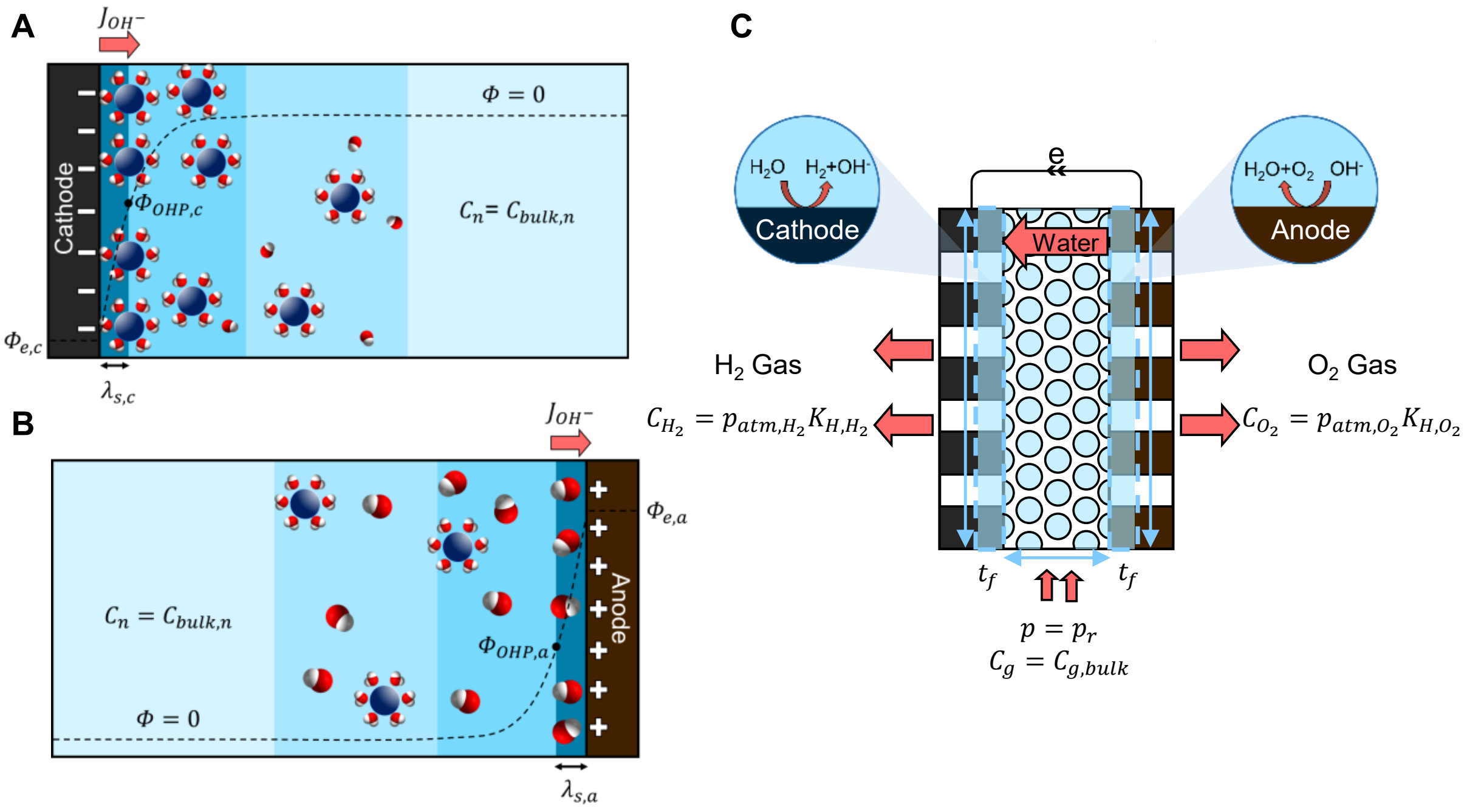

**Figure 2. Model domains and boundary conditions.** (A,B) One-dimensional cathode and anode domains used to resolve ionic transport and potential distributions across the electrode/electrolyte interfaces. (C) Two-dimensional capillary-fed electrolyzer domain used to model electrolyte delivery and dissolved $H_2$ and $O_2$ transport through the cathode, separator, and anode. The illustrations are schematic and not to scale.

### Establishing the bubble-free polarization baseline

Identically prepared electrodes were used in every cell, leaving the separator as the only component varied. The Pt/C cathode is a porous network of nanoscale catalyst aggregates, with Pt distributed uniformly across the carbon support (Figure 3A). The electrodeposited NiFeOOH anode exhibits an interconnected nanosheet morphology and homogeneous distributions of Ni and Fe (Figure 3B). These open, compositionally uniform structures provide extensive catalyst-electrolyte interfaces while maintaining pathways for electrolyte access and gas removal.

Galvanostatic electrochemical impedance spectroscopy (GEIS) was performed at 0.2 A $cm^{-2}$ to determine the area-specific series resistance of each cell (Figure 3C). The high-frequency intercepts yielded $R_s$ values of approximately 0.07, 0.06, 0.11, and 0.08 Ω $cm^2$ for the 0.45, 1.2, 5, and 8 μm separators, respectively. These values were used to calculate the corresponding $jR_s$ losses in the polarization model.

The polarization response of the 8 μm cell reproduced that reported for the original CFE configuration[17] and was therefore selected as the reference case. Experimental data below 5 mA $cm^{-2}$ were used to fit the electrode-specific exchange current densities and Tafel slopes. The resulting values were $j_{0,c} = 2.5\times10^{-4}$ A cm$^{-2}$ and $\eta_{0,c} = 30$ mV dec$^{-1}$ for the cathode, and $j_{0,a} = 1\times10^{-4}$ A cm$^{-2}$ and $\eta_{0,a} = 45$ mV dec$^{-1}$ for the anode. These parameters were then retained throughout the remaining current density range.

Two features in Figure 3D are worth noting, as we compare the experimental polarization curve to the bubble-free baseline. First, the predicted curve tracks the measurement over more than an order of magnitude in current density beyond the fitted range; the bubble-free model therefore extrapolates without additional adjustment. Second, since the model contains no bubble physics, the departure from the experiment represents bubble emergence. That this deviation current density (0.18 A cm$^{-2}$) almost coincides with the independently, acoustically determined onset (0.17 A cm$^{-2}$)[18] validates both the bubble-free baseline and the use of the deviation point as a signal of bubble nucleation. At higher current densities, the experimental voltage increases progressively above the prediction, indicating the growing influence of bubbles. This difference, $\eta_b = V_{exp} - V_{bf}$, quantifies the bubble-induced overpotential and reaches approximately 0.10 V at 1 A cm$^{-2}$.

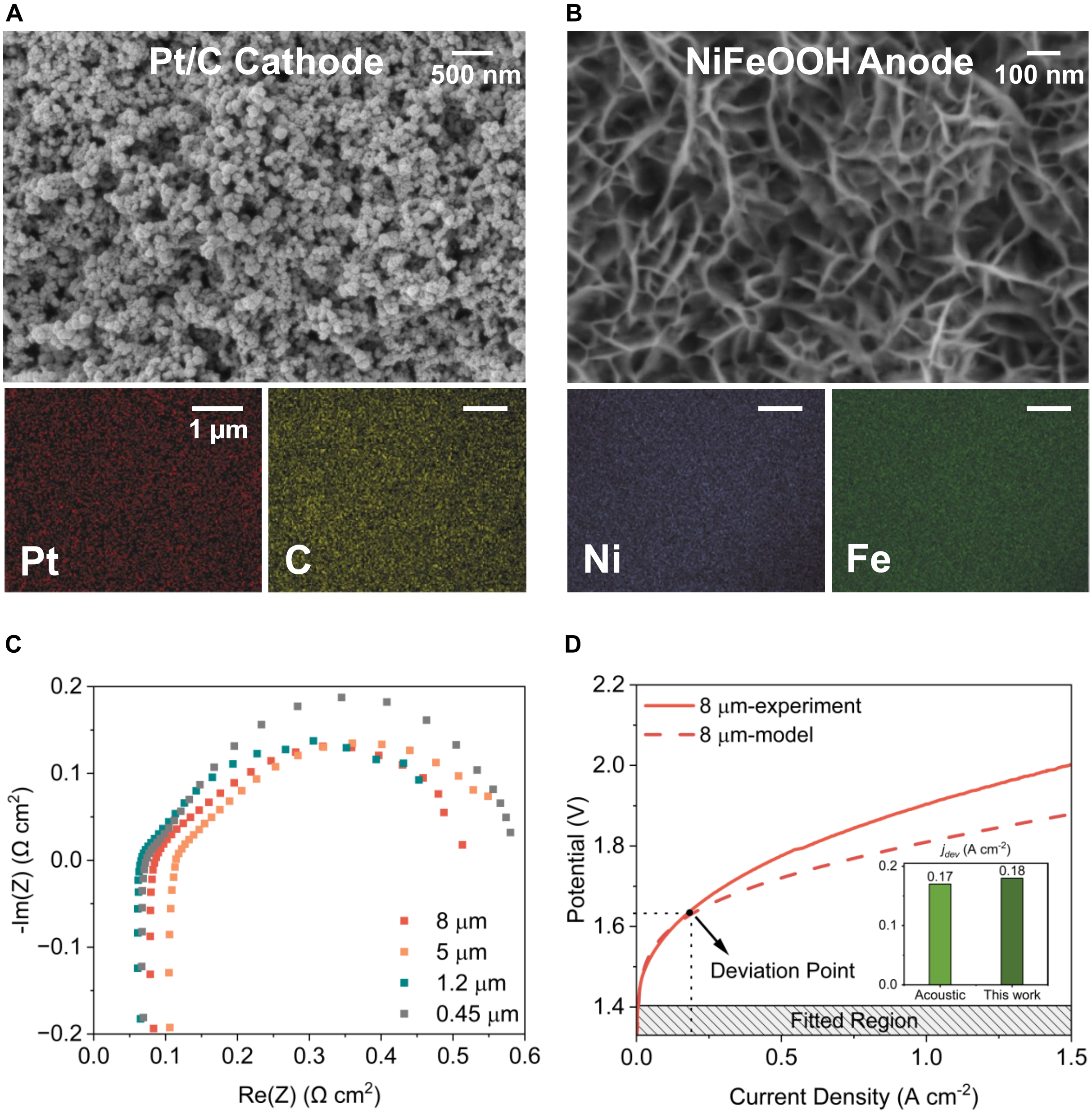

**Figure 3. Electrode characterization and validation of the bubble-free polarization baseline.** (A) SEM image of the Pt/C cathode and corresponding EDX maps of Pt and C. (B) SEM image of the NiFeOOH anode and corresponding EDX maps of Ni and Fe. (C) Nyquist plots of cells employing 0.45, 1.2, 5, and 8 μm separators. (D) Comparison of the experimental polarization curve with the bubble-free model for the 8 μm separator, showing the fitted region and identified deviation point. The inset compares $j_{dev}$ determined here with the bubble-onset current density obtained by acoustic detection.[18]

**Pore size-dependent operating limits**

Having established the bubble-free baseline, the measured cell voltages were decomposed into reversible, activation, concentration, ohmic, and bubble-induced contributions for the 8, 5, and 1.2 μm separators (Figure 4A-C). Because the same kinetic and transport parameters were used for the unchanged electrodes and electrolyte, the modeled activation and concentration overpotentials were identical across the three cells, reaching approximately 0.29 and 0.21 V, respectively, at 1 A $cm^{-2}$. The ohmic contributions were determined using the series resistances obtained from the GEIS measurements in Figure 3C and amounted to approximately 0.08, 0.11, and 0.06 V at 1 A $cm^{-2}$ for the 8, 5, and 1.2 µm separators, respectively. The higher ohmic loss of the 5 µm separator relative to the 8 µm separator is attributed to increased ionic constriction through its smaller pores. For the 1.2 µm separator, however, its higher porosity offsets this effect and provides a lower measured resistance.

At 1 A $cm^{-2}$, the 8 µm cell exhibited a bubble-induced overpotential of approximately 0.10 V and a total cell voltage of 1.90 V. Reducing the pore size to 5 µm lowered the bubble contribution to approximately 0.03 V, although its higher ohmic loss partly offset this improvement, resulting in a cell voltage of 1.88 V. For the 1.2 µm separator, the bubble-induced overpotential remained negligible throughout the investigated range up to 1.5 A $cm^{-2}$, while its ohmic loss was approximately 0.06 V, giving a cell voltage of 1.79 V at 1 A $cm^{-2}$. Overall, reducing the pore size from 8 to 1.2 µm lowered the cell voltage by approximately 0.11 V at 1 A $cm^{-2}$, of which approximately 0.09 V resulted from suppressing the bubble-induced penalty. Consistent with this reduction in bubble-induced losses, $j_{dev}$ increased from 0.18 A $cm^{-2}$ for the 8 µm separator to 0.49 and 1.73 A $cm^{-2}$ for the 5 and 1.2 µm separators, respectively. Relative to the 8 µm reference cell, the experimental bubble onset consequently increased approximately 2.7-fold for the 5 µm separator and 9.6-fold for the 1.2 µm separator.

The 0.45 and 0.1 µm separators were not included in the overpotential decomposition because their operation is governed by capillary limitation rather than bubble nucleation. As shown in Figure S5, for the 0.45 µm separator, the experimental deviation occurs at $j_{dev}$ = 1.51 A $cm^{-2}$, close to the predicted capillary limit of $j_{cap}$ = 1.85 A $cm^{-2}$ (Note S2), while the predicted nucleation limit is substantially higher at $j_{nuc}$ = 3.21 A $cm^{-2}$. This indicates that the deviation is associated with capillary limitation rather than bubble nucleation. For the 0.1 µm separator, GEIS could not be performed at 0.2 A $cm^{-2}$; therefore, $R_s$, $V_{bf}$, and $j_{dev}$ were not determined. Nevertheless, the predicted $j_{cap}$ = 0.52 A $cm^{-2}$ is substantially lower than $j_{nuc}$ = 14.43 A $cm^{-2}$, indicating capillary-controlled operation.

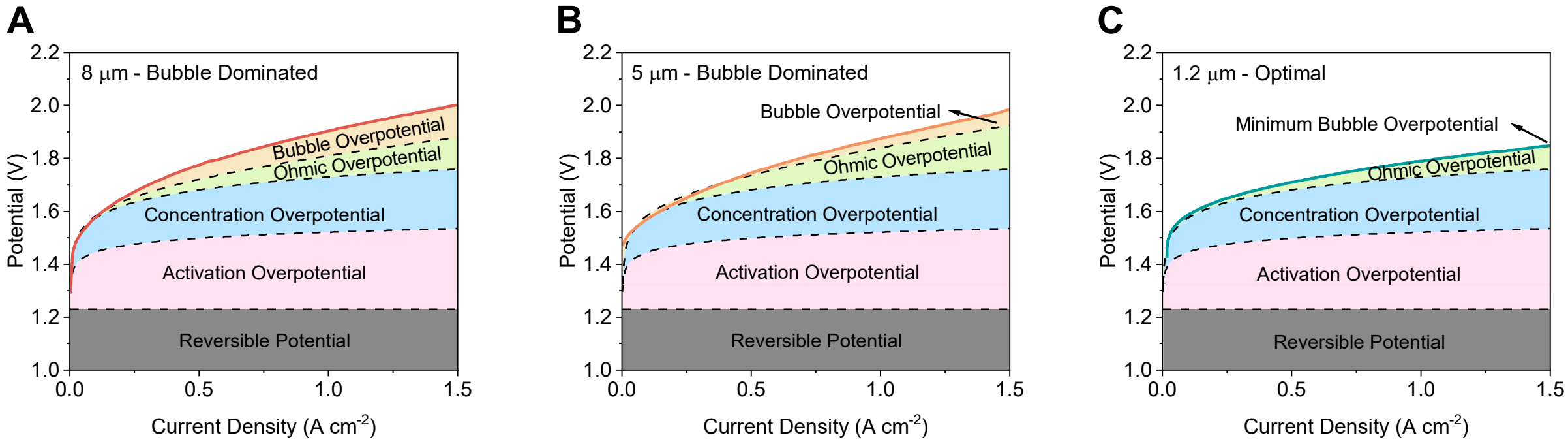

**Figure 4. Pore-size-dependent electrochemical losses.** Decomposition of the cell potential into reversible potential and activation, concentration, ohmic, and bubble-induced overpotentials for the (A) 8, (B) 5, and (C) 1.2 μm separators.

Figure 5A compares the polarization curves of different pore sizes. At 1.5 A $cm^{-2}$, the cell voltage decreased from approximately 2.00 V for the 8 μm separator to 1.98 and 1.85 V for the 5 and 1.2 μm separators, corresponding to total overpotentials of approximately 0.77, 0.75, and 0.62 V, respectively. The 0.45 μm separator initially followed the improved polarization response of the 1.2 μm separator but exhibited an increasing voltage penalty at higher current density as electrolyte replenishment approached its capillary limit. Further reducing the pore size to 0.1 μm severely restricted operation, with the current density remaining below approximately 0.05 A $cm^{-2}$ even at 2.2 V. These results indicate that decreasing pore size improves bubble suppression only while sufficient capillary electrolyte supply is maintained.

To predict $j_{nuc}$, the gas-specific nucleation factors were first determined from the 8 μm reference cell at its experimental onset of 0.18 A $cm^{-2}$. The predicted maximum dissolved gas concentrations at this condition yielded $NF_{H2} = 54.34$ and $NF_{O2} = 77.12$. Once established, these dimensionless factors serve as criteria relating the dissolved gas concentration required for nucleation to separator pore size. Applying the same values to the remaining separators, and taking the lower of the predictions for hydrogen and oxygen, gave $j_{nuc}$ values of approximately 0.29, 1.20, 3.21, and 14.43 A $cm^{-2}$ for the 5, 1.2, 0.45, and 0.1 μm separators, respectively. Thus, $j_{nuc}$ increased strongly with decreasing pore size, indicating progressively greater suppression of bubble nucleation. For the 5 and 1.2 μm separators, $j_{nuc}$ predictions are somewhat lower than the experimental $j_{dev}$ values of 0.49 and 1.73 A $cm^{-2}$, respectively, but capture the strong increase in bubble suppression with decreasing pore size. Potential sources of experimental and modeling uncertainty are discussed in Note S4.

Figure 5B reveals the central physical trade-off governing bubble-free operation. Decreasing separator pore size raises $j_{nuc}$ because the increased capillary pressure suppresses bubble nucleation. However, below an intermediate pore size, the accompanying loss of permeability outweighs the increase in capillary suction and causes $j_{cap}$ to decrease. The predicted bubble-free operating limit, $j_{lim}$, is determined by the lower of $j_{nuc}$ and $j_{cap}$. Bubble nucleation therefore controls $j_{lim}$ for the 8, 5, and 1.2 μm separators, whereas capillary replenishment controls the 0.45 and 0.1 μm separators. The experimentally observed limits follow this transition between controlling mechanisms. Consequently, the maximum attainable bubble-free current density is expected near the crossover between $j_{nuc}$ and $j_{cap}$, which lies between 0.45 and 1.2 μm for the present cell. The numerical limits reported here are specific to the present cell design, but the framework is transferable to other geometries and material properties. It can identify whether bubble nucleation or capillary limitation controls the operating window and guide optimization of parameters such as separator pore size and transport distance.

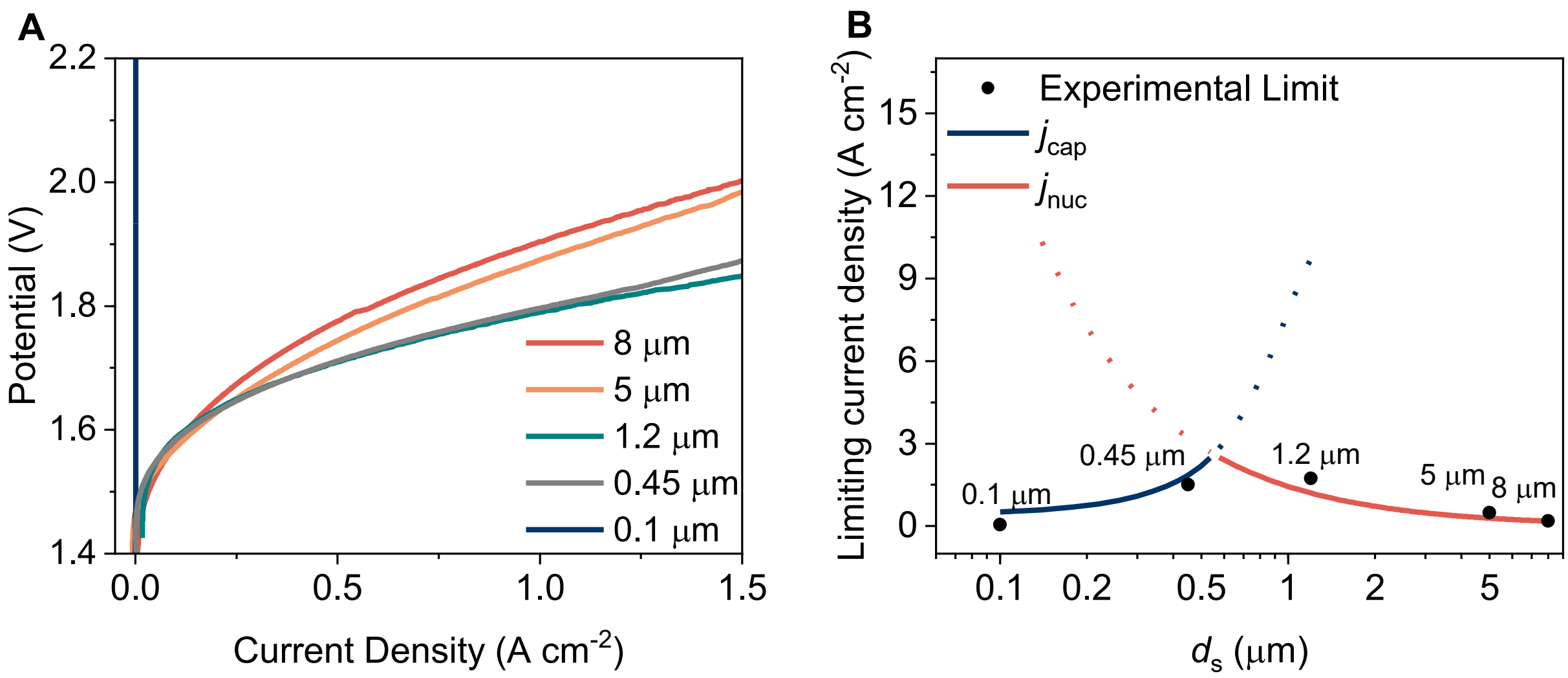

**Figure 5. Pore-size-dependent electrochemical performance and operating limits.** (A) Polarization curves of capillary-fed electrolyzers using PES separators with pore sizes of 8, 5, 1.2, 0.45, and 0.1 μm. (B) Comparison of the predicted capillary limit, $j_{cap}$, and nucleation limit, $j_{nuc}$, as a function of separator pore size. Black symbols represent $j_{dev}$ for the 8, 5, 1.2, and 0.45 μm separators and the maximum experimentally reached current density at 2.2 V for the 0.1 μm separator.

## Discussion

The results show that the bubble-free operating limit is not a fixed property of the capillary-fed architecture but emerges from two competing pore-scale constraints. Decreasing separator pore size increases the capillary barrier to gas nucleation and therefore raises $j_{nuc}$. At the same time, smaller pores reduce separator permeability and eventually lower $j_{cap}$. The attainable bubble-free current density is consequently governed by the lower of these two limits. This explains the transition from nucleation-controlled operation for the 8, 5, and 1.2 μm separators to capillary-controlled operation for the 0.45 and 0.1 μm separators.

An important implication is that a departure from the bubble-free polarization baseline should not automatically be interpreted as bubble emergence. For the 8 μm reference separator, the agreement between $j_{dev}$ = 0.18 A cm$^{-2}$ and the independently measured acoustic onset of approximately 0.17 A cm$^{-2}$ supports this interpretation in the nucleation-controlled regime. The corresponding increases in $j_{dev}$ for the 5 and 1.2 μm separators are also captured by the dissolved-gas transport model. In contrast, the deviation observed for the 0.45 μm separator is closer to its predicted capillary limit than to its nucleation limit. Combining $j_{dev}$ with the predicted $j_{nuc}$ and $j_{cap}$ therefore allows the physical origin of the experimental operating limit to be identified.

This distinction is also important for interpreting the voltage improvement. Reducing the pore size from 8 to 1.2 μm increased the bubble-onset current density from 0.18 to 1.73 A cm$^{-2}$ and reduced the bubble-induced voltage penalty to a negligible level over the investigated range. The resulting cell voltage decreased by 0.11 V at 1 A cm$^{-2}$, with most of this improvement arising from bubble suppression rather than changes in electrode kinetics. Further reduction in pore size did not provide an additional benefit because the loss of electrolyte-transport capacity became dominant. The 1.2 μm separator therefore provided the best balance among the separators investigated, rather than representing a universally optimal pore size.

More broadly, the crossover between $j_{nuc}$ and $j_{cap}$ provides a physical basis for separator design. Its position can be shifted by modifying separator porosity, permeability, wettability, thickness, electrolyte properties, or the liquid-transport distance. The framework can therefore be applied to other cell geometries and material systems to determine whether bubble nucleation or electrolyte replenishment controls operation and to identify modifications that expand the bubble-free operating window. In this way, bubble-free performance becomes a predictable and designable limit rather than an empirical threshold identified only after cell testing.

**Resource availability**

**Lead contact**

Further information and requests should be directed to the lead contact, Zhengmao Lu (zhengmao.lu@epfl.ch).

**Materials availability**

This study did not generate new unique reagents or materials that are subject to restrictions on availability.

**Data and code availability**

The data supporting the findings of this study are provided in the article and Supplemental Information. Additional data and the code used for the numerical models are available from the lead contact upon reasonable request.

**Acknowledgments**

This work was supported by the Swiss National Science Foundation (SNSF) under grant 515188.

**Author contributions**

F.K.: Conceptualization, Methodology, Investigation, Software, Formal analysis, Data curation, Visualization, Writing – original draft. Z.L.: Conceptualization, Methodology, Supervision, Funding acquisition, Writing – review & editing.

**Declaration of interests**

The authors declare no competing interests.

**Declaration of generative AI and AI-assisted technologies in the writing process**

During the preparation of this work, the authors used ChatGPT (OpenAI) to assist with improving the clarity and language of the manuscript. The authors reviewed and edited all AI-assisted content and take full responsibility for the content of the publication.

## Methods

### Materials

Ni mesh (200 LPI, 50 μm wire diameter, and 75 μm aperture; Wirescreen Xingying), $NiCl_2 \cdot 6H_2O$ (ReagentPlus®, Sigma-Aldrich), $FeCl_2 \cdot 4H_2O$ (≥99.0%, Sigma-Aldrich), KCl (Sigma-Aldrich), HCl (37 wt% in water, Sigma-Aldrich), Sigracet™ 22BB carbon fiber paper (Quintech), 10 wt% Pt supported on Vulcan XC-72 carbon (Sigma-Aldrich), Nafion™ dispersion (5% polymer content in alcohols and water, Aladdin), polyethersulfone (PES) filters with 8, 5, 1.2, 0.45, and 0.1 μm nominal pore diameters (Sterlitech Corporation), Ni 200 sheet (Goodfellow), and KOH flakes (90%, research grade, Sigma-Aldrich) were used as received without further purification.

### Electrode preparation

The Pt/C cathode was prepared following a reported procedure.[41] The catalyst ink contained 100 mg of 10 wt% Pt supported on Vulcan XC-72, 0.8 mL of 5 wt% Nafion dispersion, 1.5 mL of deionized (DI) water, and 3 mL of isopropyl alcohol (IPA). The suspension was homogenized at 10,000 rpm for 5 min (T25, IKA), sonicated for 5 min, and then spray-coated onto the microporous layer side of a 10 mm × 10 mm Sigracet™ 22BB carbon paper using an airbrush (Talon Siphon, 0.66 mm nozzle, Paasche) to a Pt loading of 0.5 mg $cm^{-2}$.

The NiFeOOH anode was prepared by cyclic electrodeposition following a reported procedure.[42] The deposition bath contained 1 M KCl, 0.075 M $NiCl_2 \cdot 6H_2O$, and 0.025 M $FeCl_2 \cdot 4H_2O$ in DI water. Prior to deposition, the Ni mesh was sonicated in IPA for 10 min, dried, pickled in 5 M HCl for 10 min, rinsed with DI water, and dried again. Electrodeposition was performed using a 10 mm × 10 mm Ni mesh as the working electrode, an oversized Ni mesh as the counter electrode, and Ag/AgCl (3 M NaCl) as the reference electrode. The electrolyte was purged with argon prior to and throughout deposition to limit $Fe^{2+}$ oxidation. Cyclic voltammetry was conducted from -1.0 to -0.2 V (vs. Ag/AgCl) at 10 mV $s^{-1}$ and room temperature using a BioLogic SP-300 potentiostat until a total deposition charge of 17 C was reached. The electrode was then rinsed with DI water, dried, and welded to the Ni bipolar plate to minimize contact resistance.

**Capillary-fed cell fabrication and assembly**

The CFE cell (Figure S6) was fabricated based on a previously reported design.[17] The cell frames incorporated an electrolyte reservoir, a bipolar plate compartment, a gas collection chamber, and electrolyte feeding lines; the cathode frame additionally incorporated an O-ring groove. Ni 200 bipolar plates (14 mm × 14 mm × 1 mm) were perforated with 36 holes of 1 mm diameter and served as both current collectors and gas diffusion layers. Before operation, 20 mm × 10 mm PES separators were immersed in DI water for 2 h and then soaked overnight in 27 wt% KOH. The components were stacked in the order Ni bipolar plate/Pt/C cathode/PES separator/NiFeOOH anode/Ni bipolar plate and aligned using leak-tight positioning screws tightened to 25 cN·m. The electrolyte reservoir was then filled with 27 wt% KOH such that 4 mm of the separator was immersed, enabling capillary wetting of the electrode interfaces, while 2 mm extended above the electrodes, giving a vertical transport distance $L$ = 14 mm.

**Electrode characterization**

The surface morphology of the prepared electrodes was examined by field emission scanning electron microscopy (FESEM, Zeiss Merlin). The Pt/C cathode and NiFeOOH anode were characterized at accelerating voltages of 1 and 3 kV, respectively. Elemental composition and spatial distribution were further analyzed using energy dispersive X-ray spectroscopy (EDX) coupled to the same FESEM system.

**Capillary rise measurement**

Capillary rise was measured at room temperature using an imaging-based rate-of-rise method adapted from a reported procedure.[43] Initially dry PES separator strips (20 mm × 10 mm) were mounted vertically, and their lower edges were brought into contact with a DI water reservoir. The advancing liquid-front height, $h$, relative to the reservoir surface was extracted from the recorded images as a function of time (Figure S3). For each separator, $h^2$ was plotted against time (Figure S4), and the water capillary-rise coefficient, $S_w$, was obtained from the slope of a linear fit in the Washburn regime. Video post processing and conversion of $S_w$ to $S_{KOH}$ for 27 wt% KOH are described in Note S3.

**Electrochemical measurements**

Electrochemical measurements of the assembled CFE cell were performed using a BioLogic SP-300 potentiostat at room temperature. Polarization curves were acquired by linear sweep voltammetry from 1.4 to 2.2 V at a scan rate of 10 mV $s^{-1}$. GEIS was performed at a DC current of 0.2 A $cm^{-2}$ with an AC perturbation of 0.02 A $cm^{-2}$ over a frequency range from 100 kHz to 100 mHz. GEIS was not performed for the 0.1 μm separator because it did not reach this current density. The series resistance, $R_s$, was taken as the high-frequency real-axis intercept, where -Im(Z) = 0. Polarization and GEIS measurements were each repeated five times for every separator, and the reported curves and values represent the means.

**Model Implementation**

The bubble-free polarization and dissolved-gas transport models were implemented and solved using COMSOL Multiphysics 6.2.

*Bubble-free polarization model*

The electrode kinetic parameters, $j_{0,i}$ and $\eta_{0,i}$, were obtained by fitting the polarization data of the 8 μm reference cell below 5 mA $cm^{-2}$ and were held fixed at higher current densities and for the remaining separators. The deviation current density, $j_{dev}$, was defined as the first current density at which the relative voltage deviation, $(V_{exp} - V_{bf})/ V_{bf}$, exceeded 0.5%. The same criterion was applied to the 8, 5, 1.2, and 0.45 μm separators. $j_{dev}$ was not determined for the 0.1 μm separator because $R_s$ and the corresponding bubble-free baseline were unavailable.

*Dissolved gas transport and bubble-onset model*

The effective gas diffusivity was calculated as $D_{eff,g,i} = D_g\varepsilon_i/ \tau_i$, where $D_g$ is the gas diffusion coefficient, $\varepsilon_i$ is the porosity, and $\tau_i$ is the tortuosity. The tortuosity is obtained from porosity using the Millington-Quirk relation, $\tau_i = \varepsilon_i^{-1/3}$.[44] The permeability $\kappa_i$ is estimated from pore size $d_i$ and porosity $\varepsilon_i$ using the Kozeny-Carman relation:[45]

$$\kappa_i = \frac{d_i^2}{180}\frac{\varepsilon_i^3}{(1-\varepsilon_i)^2}. \quad (13)$$

Water consumption and generation were represented by $Q_{m,c} = -jM_w/Ft_f$ and $Q_{m,a} = jM_w/2Ft_f$ , respectively, with $M_w$ being the molar mass of water. The corresponding hydrogen and oxygen generation rates were $R_{H2,c} = j/2Ft_f$ and $R_{O2,a} = j/4Ft_f$. All other material properties and parameters are provided in Table S1.

# Supplemental Information

# Predicting and extending the bubble-free operating limit of capillary-fed water electrolysis

*Farkhondeh Khodabandeh*[1]*, Zhengmao Lu*[1,*]

[1] Institute of Mechanical Engineering, EPFL, 1015 Lausanne, Switzerland

[*] Lead contact and correspondence: Zhengmao Lu (zhengmao.lu@epfl.ch)

### Note S1. Wetted-layer thickness sensitivity

A nominal wetted-layer thickness of $t_f$ = 2 µm was used for the gas-transport calculations presented in the main text. To evaluate sensitivity to the assumed wetted-layer thickness, $t_f$ was varied from 1 to 10 µm. At each value, the gas-specific nucleation factors were recalibrated using the 8 µm reference separator and then applied to the 5 and 1.2 µm separators. The predicted $j_{nuc}$ remained effectively unchanged as shown in Figure S1, confirming that the predictions are insensitive to $t_f$ when calibration and prediction are performed consistently.

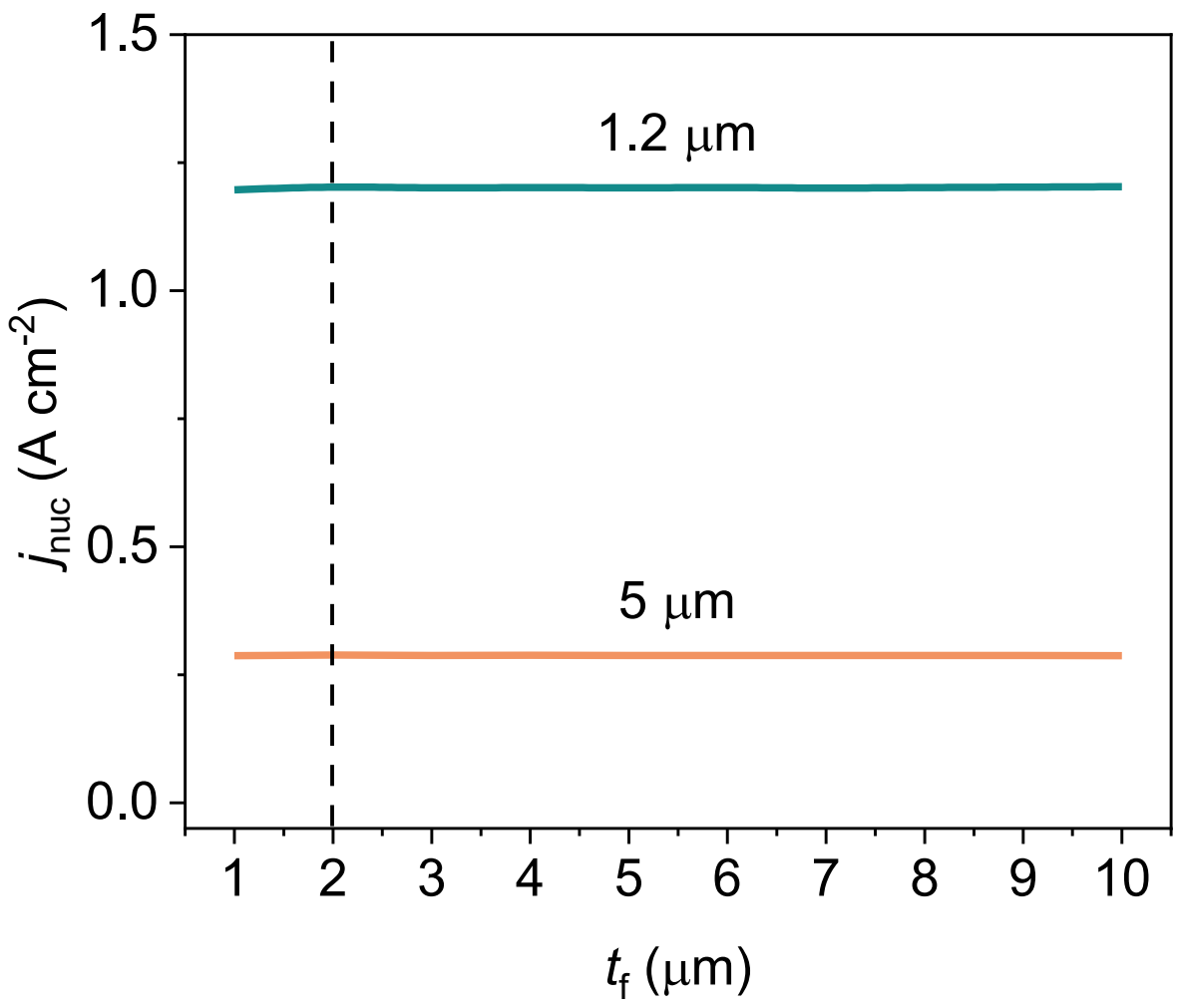

**Figure S1. Sensitivity of the predicted $j_{nuc}$ to the assumed wetted-layer thickness for the 5 and 1.2 µm separators.** The dashed line indicates the nominal value $t_f$ = 2 µm used in this work.

## Note S2. Capillary replenishment and dryout model

The cell dries out when electrochemical water consumption outpaces the maximum electrolyte supply that capillary suction can sustain. To predict the capillary-limited current density $j_{cap}$, we model electrolyte wicking vertically up the separator, over the vertical distance $L$ from the reservoir surface to the upper edge of the active electrode, through a separator of width $W$ and thickness $\delta$.

The separator's capillary transport is characterized by its capillary-rise coefficient, measured for each pore size from the rise height, $h$, plotted as $h^2$ versus time (Methods, Figure S4), using the porous-medium Washburn relation.[1] This coefficient relates the permeability $\kappa_s$ and the capillary pressure $p_{cap}$,

$$S_w = \frac{\kappa_s p_{cap}}{\varepsilon_s \mu_w}. \tag{S1}$$

where $\varepsilon_s$ is the separator porosity and $\mu_w$ is the viscosity of water. The water coefficient $S_w$ is then converted to the KOH value $S_{KOH}$ by substituting the surface tension, contact angle, and viscosity of 27 wt% KOH for those of water (Note S3).

At the capillary limit, capillary pressure balances hydrostatic head and viscous losses $\Delta p_{vis}$:

$$p_{cap} = \rho_{KOH} g L + \Delta p_{vis}, \tag{S2}$$

The viscous pressure loss and capillary pressure are evaluated as

$$\Delta p_{vis} = \frac{\mu_{KOH} Q_{max} L}{2 \kappa_s W \delta} \tag{S3}$$

$$p_{cap} = \frac{4 \gamma_{KOH} cos\theta_{KOH}}{d_s} \tag{S4}$$

where $\rho_{KOH}$ is KOH density, $g$ is gravitational acceleration, $\mu_{KOH}$ is KOH viscosity, $Q_{max}$ is the maximum electrolyte flow rate, $\gamma_{KOH}$ and $\theta_{KOH}$ are the KOH surface tension and KOH/PES contact angle, respectively, and $d_s$ is the separator pore diameter.

Because water is withdrawn to the electrode continuously along the path, the local flux decreases from its maximum at the reservoir to zero at the upper edge, so the integrated viscous loss corresponds to an effective transport length of $L/2$, giving the factor of one half in Eq. S3. A lumped constant-flux treatment, as used previously,[2] carries the full flux over the entire length $L$ and overestimates the viscous loss by a factor of two, correspondingly underpredicting the capillary limit.

The corrected maximum electrolyte flow rate, $Q_{max}$, is then

$$Q_{\mathrm{max}} = \frac{2\varepsilon_{\mathrm{s}} S_{\mathrm{KOH}} W \delta}{L}\left(1 - \frac{\rho_{\mathrm{KOH}} g L}{p_{\mathrm{cap}}}\right). \tag{S5}$$

Equating the maximum electrolyte supply to the net water consumption gives the capillary-limited current density:

$$j_{\mathrm{cap}} = \frac{2F Q_{\mathrm{max}}}{V_{\mathrm{m}} A_{\mathrm{e}}}, \tag{S6}$$

where $F$ is the Faraday constant, $V_m$ is the molar volume of water, and $A_e$ is the active geometric electrode area. All parameters used in this model are provided in Table S1.

The predicted $j_{cap}$ limits depended strongly on both separator height and pore size (Figure S2). Increasing the height lengthens the electrolyte transport pathway and increases the hydrostatic pressure loss, causing $j_{cap}$ to decrease sharply for every separator. The 5 µm separator provided the greatest capillary supply margin, followed by the 8 µm separator, consistent with their measured capillary rise coefficients. Below 5 µm, the reduction in permeability increasingly outweighed the larger capillary pressure, substantially decreasing the electrolyte replenishment capacity. At the experimental transport distance of $L$ = 14 mm, the predicted $j_{cap}$ values were approximately 22.33, 26.65, 9.73, 1.85, and 0.52 A cm$^{-2}$ for the 8, 5, 1.2, 0.45, and 0.1 µm separators, respectively. Capillary supply therefore remained well above the investigated current-density range for the 8, 5, and 1.2 µm separators, whereas the 0.45 µm separator approached its capillary limit within the experimental range and the 0.1 µm separator was strongly constrained by electrolyte replenishment.

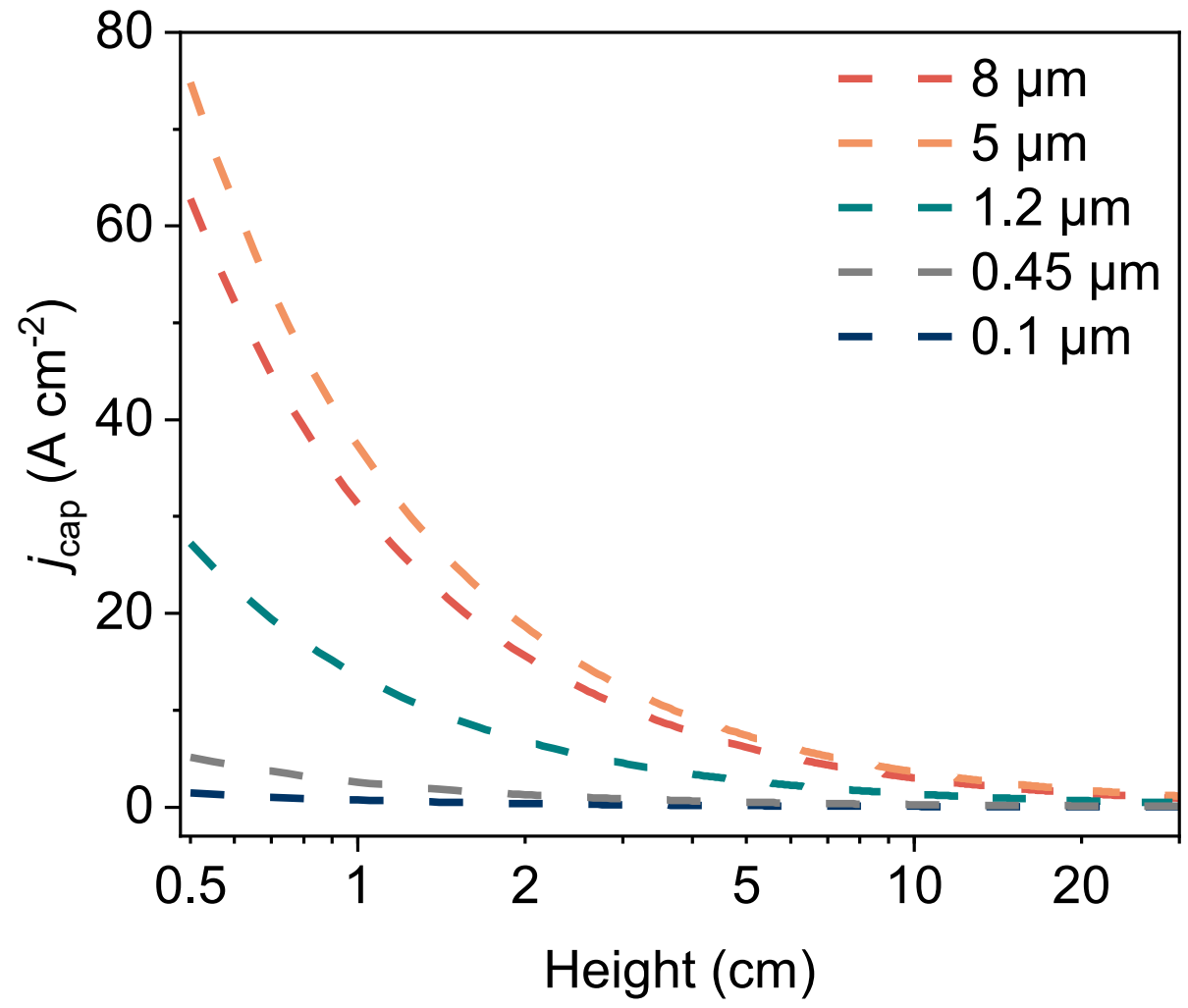

**Figure S2. Dependence of the capillary limit on separator height.** Predicted capillary-limited current density, $j_{cap}$, as a function of separator height for PES separators with pore sizes of 8, 5, 1.2, 0.45, and 0.1 μm.

### Note S3. Capillary-rise video processing and coefficient determination

The recorded videos were analyzed frame by frame using an image-based rate-of-rise method adapted from Zhang et al.[1] A region spanning the separator width was converted to grayscale and averaged laterally, and the advancing liquid front was identified from the peak in the vertical intensity gradient. Its position relative to the reservoir surface was converted to the rise height *h* using the image calibration. The water capillary rise coefficient, $S_w$, was obtained from

$$h^2 = S_w t + b, \tag{S7}$$

where *t* is time and b is the fitted intercept. The corresponding coefficient for 27 wt% KOH was calculated as

$$S_{KOH} = S_w \frac{\gamma_{KOH} cos\theta_{KOH} \mu_w}{\gamma_w cos\theta_w \mu_{KOH}}, \tag{S8}$$

where $\gamma$, $\theta$, and $\mu$ are the surface tension, contact angle, and viscosity, respectively. The subscripts KOH and w denote 27 wt% KOH and water. The parameters used are provided in Table S1.

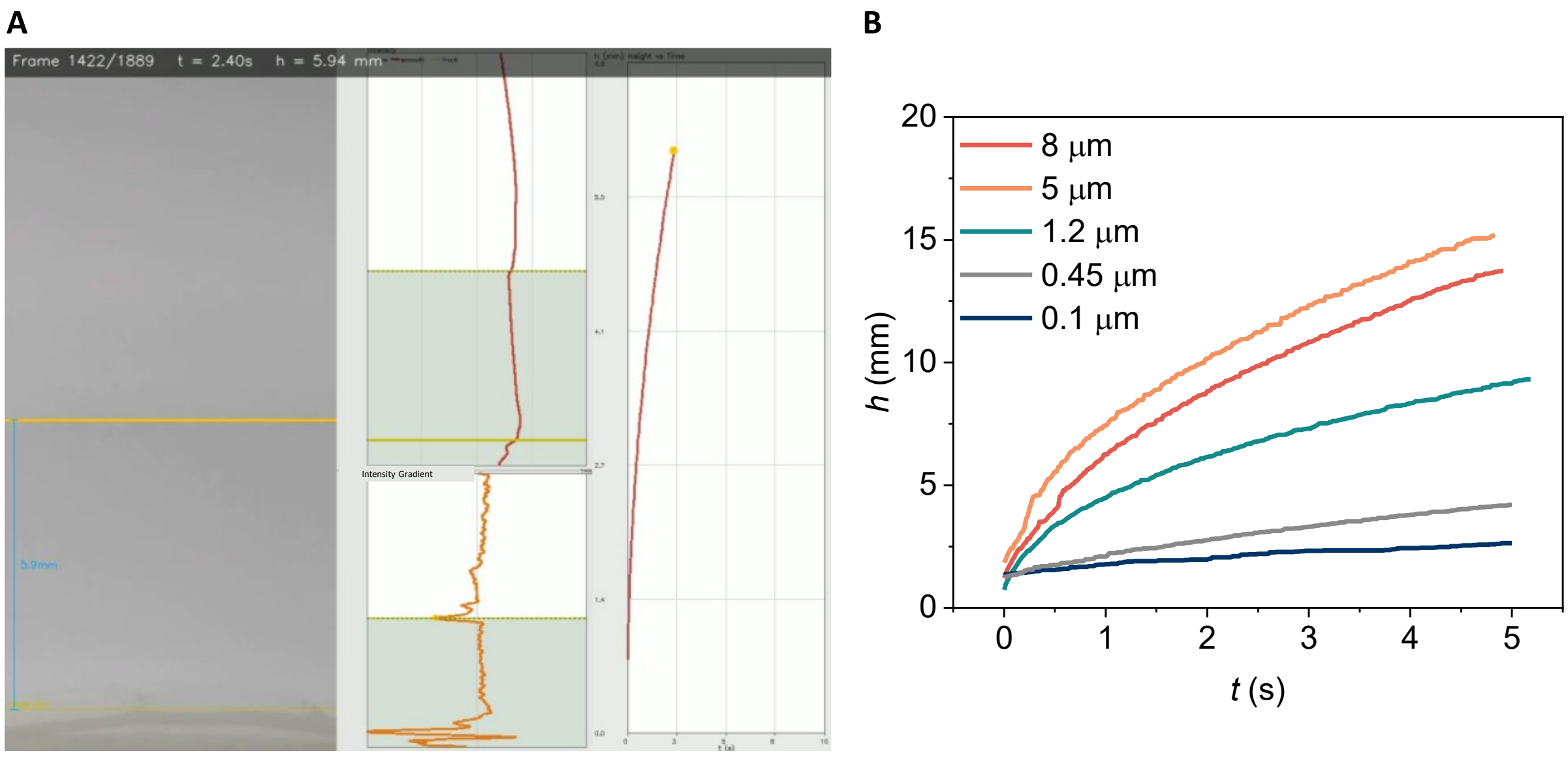

**Figure S3. Capillary rise measurement and image processing.** (A) Representative recorded frame and intensity-gradient analysis used to locate the advancing liquid front. (B) Representative rise height, *h*, as a function of time for the five separator pore sizes.

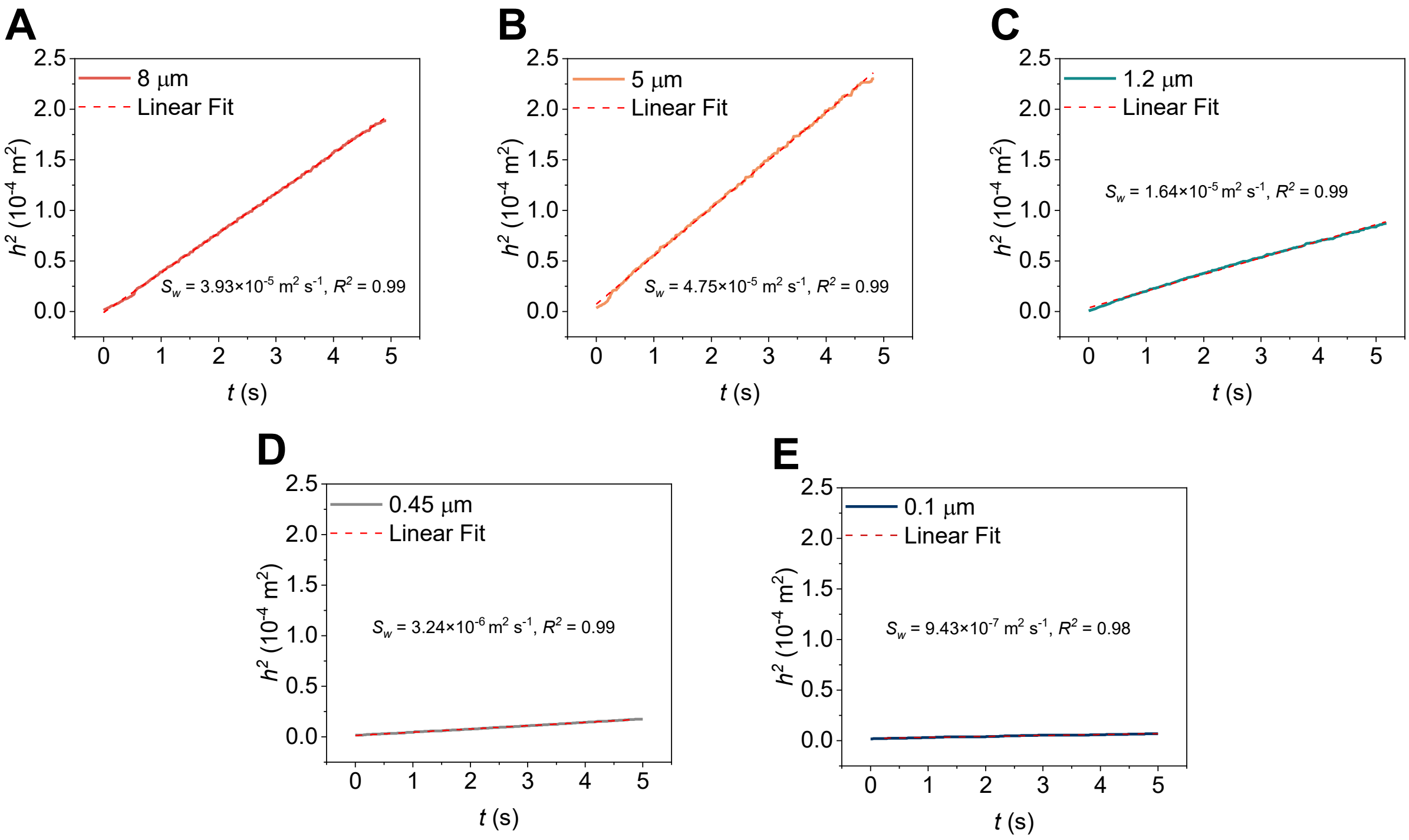

**Figure S4. Determination of the water capillary rise coefficients.** $h^2$ versus time and linear fits for PES separators with pore sizes of (A) 8, (B) 5, (C) 1.2, (D) 0.45, and (E) 0.1 µm. The fitted slopes correspond to $S_w$.

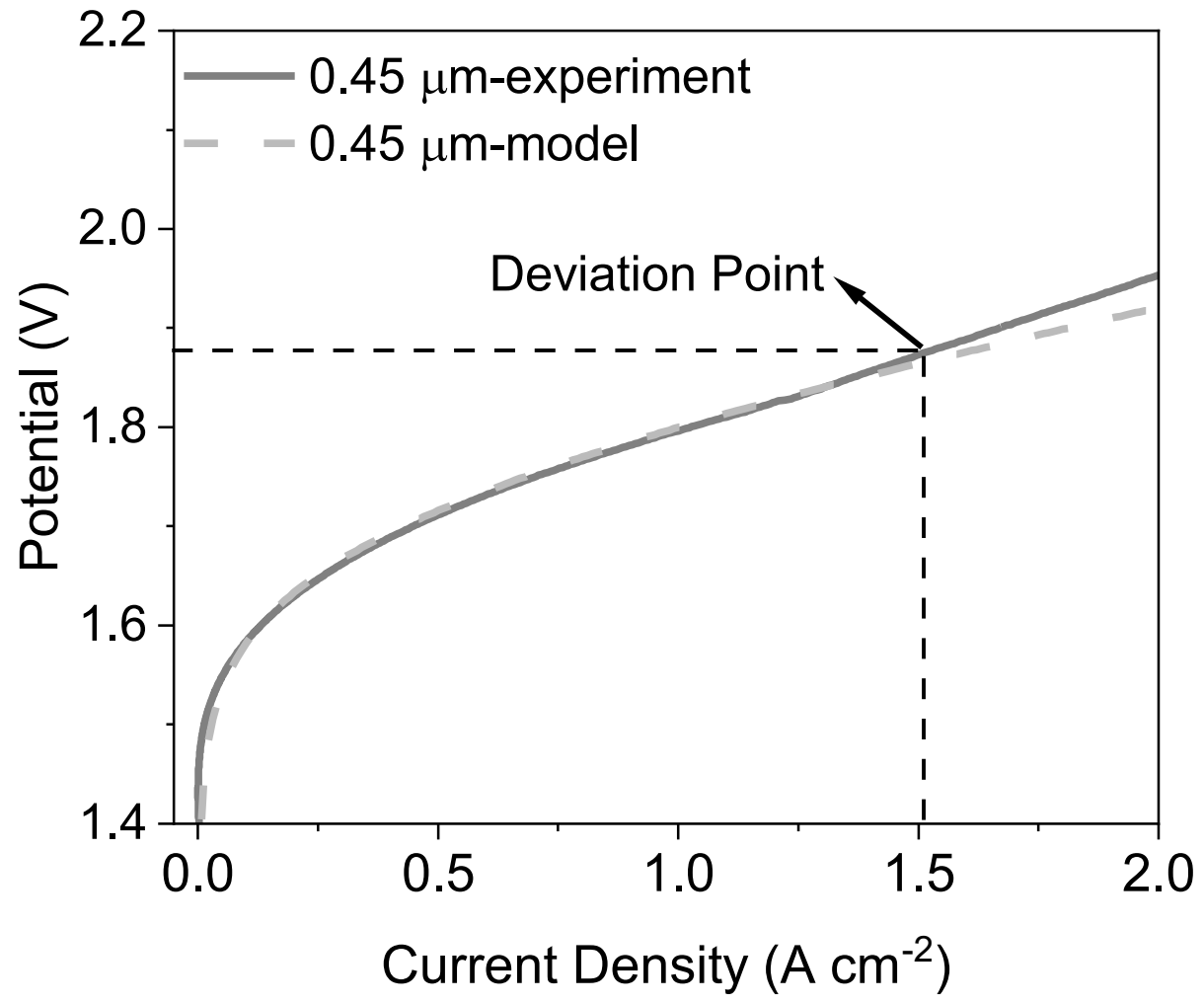

**Figure S5. Comparison of the experimental polarization curve and bubble-free model for the 0.45 µm separator.** The experimental deviation occurs at $j_{dev}$ = 1.51 A cm$^{-2}$, close to the predicted capillary limit $j_{cap}$ = 1.85 A cm$^{-2}$, while the predicted nucleation limit is substantially higher at $j_{nuc}$ = 3.21 A cm$^{-2}$. This correspondence indicates that the deviation is associated with capillary limitation rather than bubble nucleation.

**Note S4. Sources of experimental and modeling uncertainty**

Quantitative differences between the experimental and predicted operating limits can arise because $j_{dev}$ and $j_{nuc}$ do not represent exactly the same physical event. The predicted $j_{nuc}$ corresponds to the current density at which the dissolved-gas concentration first reaches the nucleation criterion, whereas $j_{dev}$ is identified from a measurable 0.5% deviation of the experimental voltage from the bubble-free baseline. Following nucleation, bubbles may need to grow and accumulate sufficiently before affecting the cell voltage, resulting in $j_{dev} > j_{nuc}$. Conversely, local loss of wetting or changes in electrical contact may produce an earlier voltage deviation unrelated to bubble nucleation. The determination of $j_{dev}$ is also affected by experimental data spacing, voltage noise, and uncertainties in the fitted bubble-free baseline. Sample-to-sample variations in separator porosity, thickness, compression, and local wetting may introduce additional variability between cells.

The gas-transport model represents each separator using a single characteristic pore size and average transport properties, whereas the actual separators contain distributions of interconnected pores with spatially varying geometry and surface characteristics. The manufacturer-reported nominal pore sizes were used in the calculations; the actual pore sizes and their distributions may differ from these values, affecting the predicted permeability, capillary pressure, and nucleation threshold. The nucleation factor, $NF_g$, was calibrated using the 8 μm reference separator and assumed to remain independent of pore size. In practice, nucleation may also depend on local pore geometry, surface roughness, wettability, and the availability of nucleation sites. Furthermore, the model assumes steady-state transport through uniformly wetted layers and does not resolve stochastic nucleation, bubble growth, or subsequent two-phase displacement. Although the sensitivity analysis indicates that the predicted $j_{nuc}$ is insensitive to the nominal wetted-layer thickness when calibration and prediction are performed consistently, spatial variations in film thickness remain unaccounted for.

Uncertainty in $j_{cap}$ arises primarily from the measured capillary-rise coefficient and its conversion from water to 27 wt% KOH. Variations in front identification, separator samples, contact angle, surface tension, and viscosity propagate into the predicted capillary limit. The capillary model also assumes complete separator wetting, uniformly distributed water withdrawal, and unchanged pore structure during cell assembly, while neglecting local gas blockage and evaporation. These effects are expected to be most important for the smallest-pore separators.

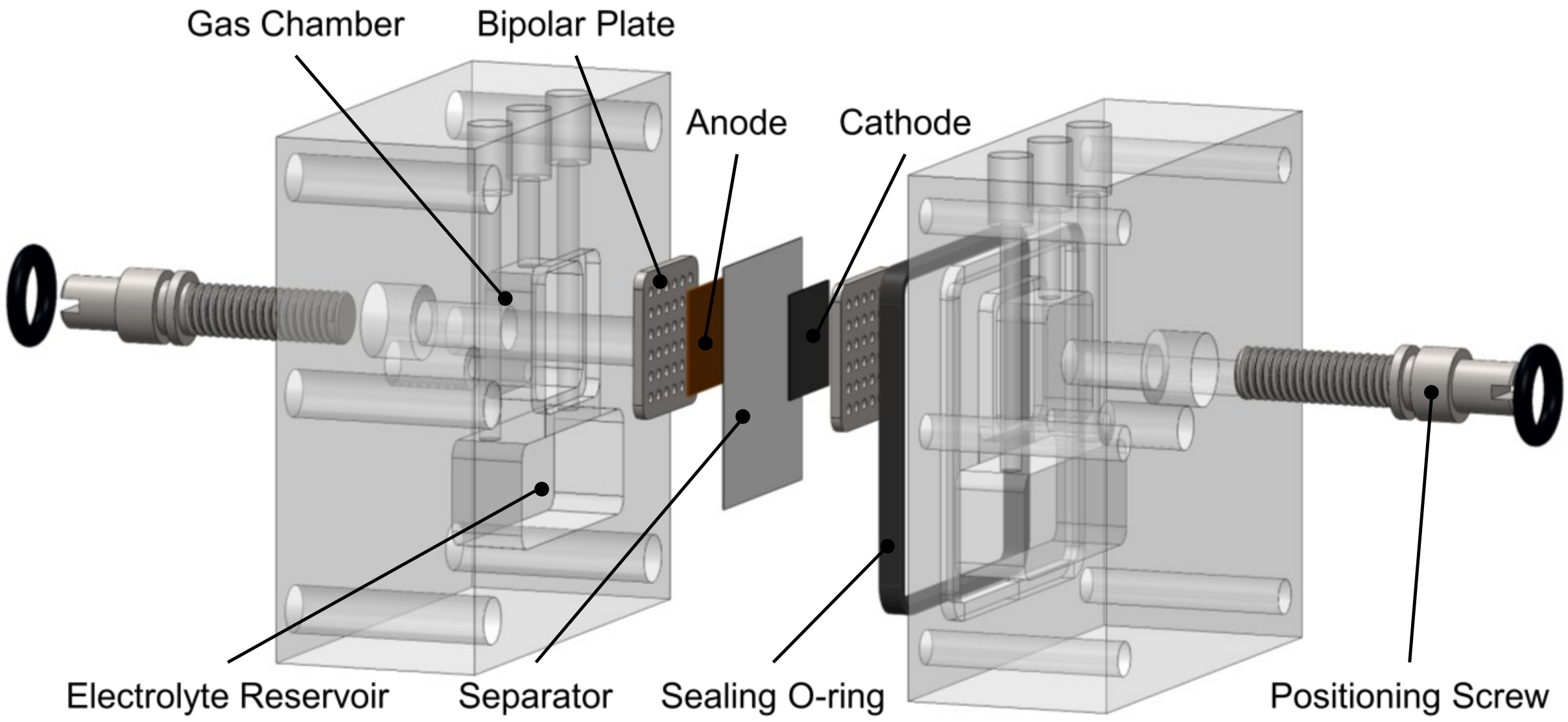

**Figure S6. Exploded view of the capillary-fed electrolyzer showing the electrolyte reservoir, gas chambers, bipolar plates, electrodes, separator, sealing O-ring, and positioning screws.**

**Table S1. Physical and transport parameters used in the polarization, dissolved gas transport, and capillary replenishment models.**

| Symbol | Value | Physical meaning | Reference |
|---|---|---|---|
| $d_c$ | 14 µm | Cathode Pore Size | - |
| $d_s$ | 8, 5, 1.2, 0.45, 0.1 µm | Separator Pore Size | - |
| $d_a$ | 75 µm | Anode Pore Size | - |
| $\delta$ | 140 µm | Separator Thickness | - |
| $L$ | 14 mm | Vertical Transport Distance | - |
| $W$ | 10 mm | Separator Width | - |
| $A_e$ | 100 $mm^2$ | Active Geometric Electrode Area | |
| $\varepsilon_c$ | 80% | Cathode Porosity | [2] |
| $\varepsilon_s$ | 8 µm: 81.3% | Separator Porosity | [2] |
| | 5 µm: 80.1% | | |
| | 1.2 µm: 84.4% | | |
| | 0.45 µm: 81.0% | | |
| | 0.1 µm: 78.2% | | |
| $\varepsilon_a$ | 36% | Anode Porosity | - |
| $\lambda_{s,c}$, $\lambda_{s,a}$ | 0.5 nm | Stern Layer Thickness | [3] |
| $z_{OH-}$ | -1 | $OH^-$ Ionic Charge Number | - |
| $z_{K+}$ | 1 | $K^+$ Ionic Charge Number | - |
| $z_{H+}$ | 1 | $H^+$ Ionic Charge Number | - |
| $a_{OH-}$ | 0.6 nm | $OH^-$ Effective Solvated Diameter | [4] |
| $a_{K+}$ | 0.662 nm | $K^+$ Effective Solvated Diameter | [4] |
| $a_{H+}$ | 0.56 nm | $H^+$ Effective Solvated Diameter | [4] |
| $C_{bulk,OH-}$ | 6.2 M | $OH^-$ Bulk Concentration | - |
| $C_{bulk,K+}$ | 6.2 M | $K^+$ Bulk Concentration | - |
| $C_{bulk,H+}$ | $10^{-9}$ M | $H^+$ Bulk Concentration | [4] |
| $C_{bulk,H2}$ | 0.5 mM | $H_2$ Bulk Concentration | [5] |
| $C_{bulk,O2}$ | 0.8 mM | $O_2$ Bulk Concentration | [5] |
| $p_{atm,H2}$ | 0.056 Pa | $H_2$ Partial Pressure | - |
| $p_{atm,O2}$ | 21.23 kPa | $O_2$ Partial Pressure | - |
| $K_{H,H2}$ | $0.79\times10^{-5}$ mol $Pa^{-1}$ $m^{-3}$ | $H_2$ Henry's Solubility Constant | [6] |
| $K_{H,O2}$ | $1.5\times10^{-5}$ mol $Pa^{-1}$ $m^{-3}$ | $O_2$ Henry's Solubility Constant | [5] |
| $D_{OH-}$ | $5.273\times10^{-9}$ m$^2$ s$^{-1}$ | $OH^-$ Diffusion Coefficient | [4] |
| $D_{K+}$ | $1.957\times10^{-9}$ m$^2$ s$^{-1}$ | $K^+$ Diffusion Coefficient | [4] |
| $D_{H+}$ | $9.311\times10^{-9}$ m$^2$ s$^{-1}$ | $H^+$ Diffusion Coefficient | [4] |
| $D_{H2}$ | $3\times10^{-9}$ m$^2$ s$^{-1}$ | $H_2$ Diffusion Coefficient | [6] |
| $D_{O2}$ | $1.5\times10^{-9}$ m$^2$ s$^{-1}$ | $O_2$ Diffusion Coefficient | [5] |
| $p_{atm}$ | 101325 Pa | Atmospheric Pressure | - |
| $\varepsilon_0$ | $8.854\times10^{-12}$ F m$^{-1}$ | Vacuum Permittivity | - |
| $\varepsilon_r$ | 78.5 | Relative Permittivity | [6] |
| $M_w$ | 18 g mol$^{-1}$ | Water Molar Mass | - |
| $V_m$ | 18 mL mol$^{-1}$ | Water Molar Volume | - |
| $\rho_{KOH}$ | 1.29 g cm$^{-3}$ | KOH Density | [2] |
| $T$ | 298.15 K | Temperature | - |
| $F$ | 96485 C mol$^{-1}$ | Faraday Constant | - |
| $R$ | 8.314 J mol$^{-1}$ K$^{-1}$ | Gas Constant | - |
| $N_A$ | $6.022\times10^{23}$ mo1$^{-1}$ | Avogadro Constant | - |
| $g$ | 9.8067 m s$^{-2}$ | Gravitational Acceleration | - |
| $\gamma_{KOH}$ | 83 mN m$^{-1}$ | KOH Surface Tension | [2] |
| $\mu_{KOH}$ | 2.8 mPa s | KOH Viscosity | [2] |
| $\gamma_w$ | 72 mN m$^{-1}$ | Water Surface Tension | - |
| $\mu_w$ | 1 mPa s | Water Viscosity | - |
| $\theta_{KOH}$ | 46.97° | KOH/PES Contact Angle | [2] |
| $\theta_w$ | 31.07° | Water/PES Contact Angle | [2] |
| $S_w$ | 8 µm: $3.93 \times 10^{-5}$ m$^2$ s$^{-1}$<br>5 µm: $4.75 \times 10^{-5}$ m$^2$ s$^{-1}$<br>1.2 µm: $1.64 \times 10^{-5}$ m$^2$ s$^{-1}$<br>0.45 µm: $3.24 \times 10^{-6}$ m$^2$ s$^{-1}$<br>0.1 µm: $9.43 \times 10^{-7}$ m$^2$ s$^{-1}$ | Water Capillary-Rise Coefficient | This work |
| $S_{KOH}$ | 8 µm: $1.29 \times 10^{-5}$ m$^2$ s$^{-1}$<br>5 µm: $1.56 \times 10^{-5}$ m$^2$ s$^{-1}$<br>1.2 µm: $5.38 \times 10^{-6}$ m$^2$ s$^{-1}$<br>0.45 µm: $1.06 \times 10^{-6}$ m$^2$ s$^{-1}$<br>0.1 µm: $3.09 \times 10^{-7}$ m$^2$ s$^{-1}$ | KOH Capillary-Rise Coefficient | This work |